\documentstyle[preprint,aps]{revtex}
\begin{document}
\draft

\preprint{MIT/CTP Preprint \# 2674}
\title{ Relativistic Effects in the Electromagnetic Current
at GeV Energies}

\author{S. Jeschonnek and T. W. Donnelly}

\address{\small \sl Center for Theoretical Physics, Laboratory for
Nuclear Science and Dept. of Physics, Massachusetts Institute of
Technology, Cambridge, MA 02139, U.S.A.}

\date{\today}
\maketitle


\begin{abstract}
We employ a recent approach to the non-relativistic reduction of the
electromagnetic current operator in 
calculations of electronuclear reactions. In contrast to the
traditional scheme, where approximations are made for the transferred
momentum, transferred energy and initial momentum of the struck
nucleon in obtaining an on-shell inspired form for the current, we
treat the problem exactly for the transferred energy and transferred
momentum. We calculate response functions for the reaction $^2H(e,e'p)n$
at CEBAF (TJNAF) energies and find large relativistic corrections. We also
show that in Plane Wave Impulse Approximation, it is always possible
to use the full operator, and we present a comparison of such a
limiting case with the results incorporating relativistic effects to the
first order in the initial momentum of the struck nucleon.
\end{abstract}

\section{Introduction}

At present, there exists a broad experimental program
of electron scattering studies at TJNAF, MAMI,
Bates, and NIKHEF aimed at understanding the short range
structure of nuclei and the properties of nucleons in the nuclear
medium. 
In the theoretical calculation of these processes, one typically performs
a non-relativistic reduction of the relativistic electromagnetic
current operator, as the nuclear wave function is usually given in a
non-relativistic framework.
Traditionally, certain assumptions have been made for the
non-relativistic reduction: the momentum $\vec q$ transferred from the
electron to the nucleus is regarded as being smaller than the nucleon mass
$m_N$, and the transferred energy $\omega$ and the initial momentum
$\vec p$ of the nucleon are in turn assumed to be smaller than the
transferred momentum, namely,
\begin{eqnarray}
|\vec q| < m_N \nonumber \\
\omega \approx \frac{\vec q \, ^2}{2 m_N} \ll m_N \\
\label{nonrel}
|\vec p| \ll  m_N \nonumber \,.
\end{eqnarray}
These approximations often cannot be justified for present day experiments,
since  the transferred energies and
momenta may be in the GeV region and thus are comparable with or even
bigger than the nucleon mass. This applies especially to TJNAF with
its 4 GeV electron beam.

In the past, there have been several attempts to improve on the
reduction by expanding in powers of $q / m_N$ \cite{mcvoy,boffi}, but
for a situation where also the transferred energies and not only the
transferred momenta attain values comparable to the nucleon mass,
these approaches are insufficient.  Such a situation occurs and is
indeed a common one, for instance, in near quasielastic kinematics.
There also exist other expansion schemes \cite{aren}.

In this paper, we focus on the problem of an improved treatment of the
non-relativistic reduction of the electromagnetic current operator.
We do not discuss here, however, the general problem of off-shell
prescriptions for the current. Instead, since we treat the
non-relativistic reduction in a nuclear framework, we adopt the
popular ansatz of using the on-shell form for the current and employ
this form with bound and scattering wave functions for the nucleons in
the nucleus.  In so-doing this, we are, of course, using yet another
off-shell prescription, since only with plane waves will the current
matrix elements be taken strictly on-shell.  The goal of the paper is
to explore specific classes of non-relativistic approximations within
the context of this particular ansatz, identified in this work as the
``on-shell form'' of the off-shell current.

Some aspects of the formalism presented in this paper have been used
in other recent publications \cite{pvquique,quiqueinc,quiquecoin},
although several approximations invoked in those studies have not been
checked explicitly before using a realistic nuclear model. In this
paper, we point out how to check the quality of these approximations
and we present a comparison between the exact results (i.e. exact
within the context of the model; see below for details) and the
approximated results employing a realistic nuclear model. The
calculations shown in this paper have been carried out for the
reactions $^2H (e,e'p)n$ and $^2 \vec H (\vec e,e'p)n$ at CEBAF
energies, which are interesting in their own right.

This paper is organized as follows: First, we discuss the new approach
for the single-nucleon current that treats the problem exactly for the
transferred energy and transferred momentum.  The approach employs the
on-shell form for use in off-shell calculations. In particular, the
current is expanded in powers of the initial nucleon momentum.  We
show that for the special case of Plane Wave Impulse Approximation and
exclusive $(e,e'N)$ reactions, we can always employ the full form of
the current operator. This allows us to check the quality of the
expansion in powers of the initial nucleon momentum.  Then we compare
results which have been obtained using the non-relativistic reduction
and the improved current operator, and discuss the role of the new
contributions arising from the relativistic corrections.

\section{The Current Operator}

We start our discussion with the single-nucleon on-shell
electromagnetic current operator and its non-relativistic
reduction. Afterwards, we discuss how the current is applied in the
nucleus. Throughout this paper, we use the conventions of Bjorken and
Drell \cite{bjorkendrell} and base our notation on the abbreviated
treatment presented in \cite{pvquique}.
 
It is useful to rewrite the single-nucleon current,

\begin{equation}
J^\mu (P\Lambda ;P^{\prime }\Lambda ^{\prime })=\bar u(P^{\prime }\Lambda
^{\prime })\left[ F_1\gamma ^\mu +\frac i{2m_N}F_2\sigma ^{\mu \nu }Q_\nu
\right] u(P\Lambda ),  \label{sn1}
\end{equation}
in a form that is more suitable for application to nuclear
problems.  The 4-momentum of the incident nucleon is $P^\mu =(E,\vec
p)$, the 4-momentum of the outgoing nucleon is $P^{\prime \mu
}=(E^{\prime },\vec p \, ^{\prime })$, and the transferred 4-momentum is
$Q^\mu =P^{\prime \mu }-P^\mu $. The spin projections for incoming and
outgoing nucleons are labeled $\Lambda $ and $\Lambda ^{\prime }$ ,
respectively.

It is convenient to introduce dimensionless variables at this point: 
\begin{eqnarray}
\lambda &\equiv &\omega /2m_N  \nonumber \\
\vec{\kappa } &\equiv &\vec{q}/2m_N  \nonumber \\
\tau &\equiv &\kappa ^2-\lambda ^2  \label{sn2} \\
\vec{\eta } &\equiv &\vec{p}/m_N  \nonumber \\
\varepsilon &\equiv &E/m_N=\sqrt{1+\eta ^2}  \, . \nonumber
\end{eqnarray}
For the outgoing nucleon, $\vec \eta \,'$ and $\epsilon \, '$ are defined
correspondingly. Of course, these latter quantities can be eliminated
through 4-momentum conservation.

The Dirac and Pauli form factors are functions
only of the 4-momentum transfer, $F_{1,2}=F_{1,2}(\tau )$, and in the
following we use the Sachs form factors 
$G_E(\tau ) = F_1(\tau )-\tau F_2(\tau )$ and
$G_M(\tau ) = F_1(\tau )+F_2(\tau )$.

Also, it is useful to introduce the angle $\theta $ between
$\vec{\kappa }$ and $ \vec{\eta }$, which allows us to obtain the
following relations between the kinematic variables:
 
\begin{eqnarray}
\kappa \eta \cos \theta &=&\lambda \varepsilon -\tau 
\nonumber
\\
\tau \left( \varepsilon +\lambda \right) ^2 &=&\kappa ^2\left[ 1+\tau
+\delta ^2\right] ,  \label{sn7}
\end{eqnarray}
where $\delta$ is defined as $\delta \equiv \eta \sin \theta .$ For
later use we also define:
\begin{eqnarray}
\mu _1 &\equiv &\frac{\kappa \sqrt{1+\tau }}{\sqrt{\tau }\left( \varepsilon
+\lambda \right) }=\frac 1{\sqrt{1+\frac{\delta ^2}{1+\tau }}}  \label{sn8}
\\
\mu _2 &\equiv &\frac{2\kappa \sqrt{1+\tau }}{\sqrt{\tau }\left( 1+\tau
+\varepsilon +\lambda \right) }=\frac{2\mu _1}{1+\frac{\sqrt{\tau (1+\tau )}}%
\kappa \mu _1}.  \label{sn9}
\end{eqnarray}

It is our aim to obtain 
expressions for the single-nucleon electromagnetic
current operators $\bar{J}^{\mu }(P;P^{\prime })$ that occur inside
the two component spin-$\frac{1}{2}$ spinors, {\it viz.}
\begin{equation}
J^{\mu }(P\Lambda ;P^{\prime }\Lambda ^{\prime })\equiv \chi _{\Lambda
^{\prime }}^{\dagger }\bar{J}^{\mu }(P;P^{\prime })\chi _{\Lambda }^{{}}.
\label{sn11}
\end{equation}
The bar over the current distinguishes an operator from its spin
matrix elements. Writing these in the following way with an overall
factor $ f_{0}$ removed for convenience (note that $J^{\mu }$ in
Eq. (\ref{sn11}) is a four-vector, whereas $V^{\mu }$ in
Eq.(\ref{sn12}) is not, although also labeled with the Lorentz index
$\mu $),
\begin{eqnarray}
\bar{J}^{\mu } &\equiv &f_{0}V^{\mu }  \label{sn12} \\
f_{0} &\equiv &\frac{1}{\mu _{1}\sqrt{1+\frac{\tau }{4\left( 1+\tau \right) }%
\mu _{2}^{2}\delta ^{2}}},  \label{sn13}
\end{eqnarray}
the electromagnetic current operator may then be rewritten in terms of
the kinematical variables introduced previously:
\begin{eqnarray}
V^{0} &=&\xi _{0}+i\xi _{0}^{\prime }\left( \vec{\kappa }\times \vec{%
\eta }\right) \cdot \vec{\sigma }  \label{sn14} \\
V^{3} &=&\left( \lambda /\kappa \right) V^{0}  \label{sn15} \\
\vec{v}^{\bot } &=&\xi _{1}\left[ \vec{\eta }-\left( \frac{\vec{%
\kappa }\cdot \vec{\eta }}{\kappa ^{2}}\right) \vec{\kappa }\right] -i%
\bigg\{\xi _{1}^{\prime }\left( \vec{\kappa }\times \vec{\sigma }%
\right)   \label{sn16} \\
&&\left. +\xi _{2}^{\prime }\left( \vec{\kappa }\cdot \vec{\sigma }%
\right) \left( \vec{\kappa }\times \vec{\eta }\right) +\xi
_{3}^{\prime }\left[ \left( \vec{\kappa }\times \vec{\eta }\right)
\cdot \vec{\sigma }\right] \left[ \vec{\eta }-\left( \frac{\vec{%
\kappa }\cdot \vec{\eta }}{\kappa ^{2}}\right) \vec{\kappa }\right]
\right\} ,  \nonumber
\end{eqnarray}
where the coefficients $\xi_i $ (no spin dependence) and $\xi ^{\prime
}_i$ (spin dependence) are given by:
\begin{eqnarray}
\xi _{0} &=&\frac{\kappa }{\sqrt{\tau }}\left[ G_{E}+\frac{\mu _{1}\mu _{2}}{%
2(1+\tau )}\delta ^{2}\tau G_{M}\right]   \nonumber \\
\xi _{0}^{\prime } &=&\frac{1}{\sqrt{1+\tau }}\left[ \mu _{1}G_{M}-\frac{1}{2%
}\mu _{2}G_{E}\right]   \nonumber \\
\xi _{1} &=&\frac{1}{\sqrt{1+\tau }}\left[ \mu _{1}G_{E}+\frac{1}{2}\mu
_{2}\tau G_{M}\right]   \nonumber \\
\xi _{1}^{\prime } &=&\frac{\sqrt{\tau }}{\kappa }\left( 1-\frac{\mu _{1}\mu
_{2}}{2(1+\tau )}\delta ^{2}\right) G_{M}  \nonumber \\
\xi _{2}^{\prime } &=& \frac{\lambda \sqrt{\tau }}{2\kappa ^{3}}\mu _{1}\mu
_{2}G_{M}  \nonumber \\
\xi _{3}^{\prime } &=&\frac{\sqrt{\tau }}{2\kappa (1+\tau )}\mu _{1}\mu
_{2}\left[ G_{E}-G_{M}\right] . 
 \label{sn17}
\end{eqnarray}
These are exact expressions for the on-shell electromagnetic current
operator, also given in \cite{pvquique}.  So far, Eq. (\ref{sn1}) has
only been rewritten.  The conservation of the current is obvious from
Eq. (\ref{sn15}).

Later on, we will refer to the operator associated with $\xi_o$ as
zeroth-order charge operator, where the zeroth-order indicates that
the operator, in this case the identity, is of zeroth order in $\eta$.
Correspondingly, we call the term containing the $\xi_o'$ first-order
spin-orbit operator, the term containing $\xi_1$ first-order
convection current, the term containing $\xi_1'$ zeroth-order
magnetization current, the term containing $\xi_2'$ first-order
convective spin-orbit term, and the term containing $\xi_3'$
second-order convective spin-orbit term.  Note that all coefficients
$\xi_i$ and $\xi_i'$ contain terms that are either of zeroth or first
order in $\eta$.

Compared with the fully non-relativistic reduction discussed below
(compare Eqs. (\ref{curnr})), there are several new types of operators
in Eqs. (\ref{sn14}-\ref{sn16}), and the operators which are also
present for the non-relativistic reduction, namely the zeroth-order
charge operator, the zeroth-order magnetization current and the
first-order convection current, are multiplied with new factors.  For
the charge, we now have an additional contribution, which is
traditionally called the ``spin-orbit'' part of the charge. It is also
present in the $q / m_N$ expansion schemes.  In the transverse part of
the current, there are two new operators with different spin
structures (the first-order and second-order convective spin-orbit
terms).

Eqs. (\ref{sn14}) and (\ref{sn16}), which are coordinate free, can
also be rewritten using a coordinate system with unit vectors
$\vec{u}_3\equiv \vec{\kappa /}\kappa $, $\vec{u}_2\equiv \left(
\vec{\kappa }\times \vec{\eta }\right) /\kappa \eta \sin \theta $ and
$\vec{u}_1\equiv \vec{u}_2\times \vec{u}_3$ to obtain
\begin{eqnarray}
V^0 &=&\nu _0+i\nu _0^{\prime }\sigma ^2  \label{sn18} \\
\vec{v}^{\bot } &=&\nu _1\vec{u}_1+i\left[ \nu _1^{\prime }\sigma ^2%
\vec{u}_1-\left( \nu _2^{\prime }\sigma ^1+\nu _2^{\prime \prime
}\sigma ^3\right) \vec{u}_2\right] ,  \label{sn19}
\end{eqnarray}
with 
\begin{eqnarray}
\nu _0 &=&\frac \kappa {\sqrt{\tau }}\left[ G_E+\frac{\mu _1\mu _2}{2(1+\tau
)}\delta ^2\tau G_M\right] =\xi _0  \nonumber \\
\nu _0^{\prime } &=&\frac \kappa {\sqrt{1+\tau }}\left[ \mu _1G_M-\frac
12\mu _2G_E\right] \delta =\kappa \delta \xi _0^{\prime }  \nonumber \\
\nu _1 &=&\frac 1{\sqrt{1+\tau }}\left[ \mu _1G_E+\frac 12\mu _2\tau
G_M\right] \delta =\delta \xi _1  \nonumber \\
\nu _1^{\prime } &=&\sqrt{\tau }\left( G_M-\frac{\mu _1\mu _2}{2(1+\tau )}%
\delta ^2G_E\right) = \kappa \xi_1'-\kappa \delta^2 \xi_3'  \label{sn20} \\
\nu _2^{\prime } &=&\sqrt{\tau }\left( 1-\frac{\mu _1\mu _2}{2(1+\tau
)}\delta ^2\right) G_M=\kappa \xi _1^{\prime }  \nonumber \\
\nu _2^{\prime \prime } &=&=\frac 12\left( \frac \lambda \kappa \right) 
\sqrt{\tau }\mu _1\mu _2\delta G_M=\kappa ^2\delta \xi _2^{\prime } 
\, . \nonumber
\end{eqnarray}
In addition to expressing the current operator with respect to the
coordinate system with unit vectors $\vec u_1, \vec u_2, \vec u_3$,
the terms in Eqs. (\ref{sn18}),(\ref{sn19}) have also been reordered
according to which kind of Pauli matrix they contain, so that it is
easy to see which terms can interfere.  One can read off immediately
the orders of the terms that contribute to the cross section for electron
nucleon scattering. Additionally, here it is easy  to see which
terms do not flip the spin (those involving either no Pauli matrix or 
$\sigma^3$, namely, $\nu_o$, $\nu_1$, and $\nu_2''$) and which do
(those involving $\sigma^{1,2}$, and hence $\sigma^{\pm}$, namely, 
$\nu_o'$, $\nu_1'$, and $\nu_2'$).

It is instructive to evaluate the spin matrix elements of the operator
given in Eqs. (\ref{sn18}) and (\ref{sn19}) and from these to obtain the
unpolarized single-nucleon responses (see also \cite{adm}).  For
example, the purely longitudinal ``L'' response involves non-spin-flip
and spin-flip matrix elements of $V^o$ squared and added 
incoherently (in the unpolarized responses, the two contributions
cannot interfere):

\begin{eqnarray}
f_0^2\left[ \nu _0^2+\nu _0^{\prime 2}\right] &=&\frac 1{1+\tau
}\left\{ \left( \varepsilon +\lambda \right) ^2G_E^2+\kappa ^2\delta
^2G_M^2\right\} \nonumber \\ &=&\frac{\kappa ^2}\tau \left\{
G_E^2+\delta ^2W_2\right\} \, , \label{sn22} \\
\end{eqnarray}
where $ W_1 \equiv \tau G_M^2$ and $ W_2 \equiv \frac 1{1+\tau }\left[
G_E^2+\tau G_M^2\right]$. Similarly, the purely transverse parts of the 
current yield
\begin{eqnarray}
f_0^2\left[ \nu _1^2+\nu _1^{\prime 2}\right] &=&\frac 1{\kappa ^2}\frac
1{1+\tau }\left\{ \left( \varepsilon +\lambda \right) ^2\tau ^2G_M^2+\kappa
^2\delta ^2G_E^2\right\}  
\nonumber \\
&=&W_1+\delta ^2W_2  \label{sn24} \\
f_0^2\left[ \nu _2^{\prime 2}+\nu _2^{\prime \prime 2}\right] &=&\tau
G_M^2=W_1 \, ,  \label{sn25}
\end{eqnarray}
and from this one can see that the unpolarized ``T'' and ``TT''
responses (involving the sum and minus the difference of
Eqs.(\ref{sn24}-\ref{sn25}), respectively) are $2 W_1 + \delta^2  W_2$
and $- \delta^2 W_2$, respectively.
Finally, the unpolarized ``TL'' response is
\begin{eqnarray}
f_0^2\left[ \nu_o \nu_1 + \nu_o' \nu_1' \right ] & = &
\frac{\kappa}{\sqrt{\tau}} \sqrt{1 + \tau +\delta^2} \, \delta W_2
\, .\label{sn25a}
\end{eqnarray}
Note that in Eqs. (\ref{sn22}-\ref{sn25a}) the form factors $G_E$ and
$G_M$ enter only in linear combinations of $G_E^2$ and $G_M^2$.  Terms
of the type $G_EG_M$ do not occur, as expected.

When used in PWIA for the unpolarized cross section (including the
special case of the relativistic Fermi gas model \cite{cdm}) these
constitute the on-shell forms for the single-nucleon electromagnetic
response functions to be employed in concert with the nuclear spectral
function. They bear a strong resemblance to the popular off-shell
prescriptions that are widely used in treatments of $(e,e'N)$
reactions \cite{deforest,koch,cdp} and, in fact, at least in
reasonably ``safe'' situations such as nearly quasifree kinematics
yield results that do not differ appreciably from those of the latter.
Of course, our ultimate goal is to provide current operators and their
matrix elements taken with interacting initial {\em and} final state
wave functions, in which case Eqs.(\ref{sn22}-\ref{sn25a}) do not hold
and, of course, the cross section does not factorize into
single-nucleon responses multiplied by a spectral function.

In electronuclear reactions, one usually assumes that the electron
emits one photon which interacts with one of the nucleons in the
nucleus. Nucleons in the nucleus are bound and therefore off-shell;
they do not fulfill the same energy-momentum relations as free
nucleons.  Currently, there exists no microscopic description of this
off-shell behavior that can be applied for a wide range of kinematic
conditions --- there are only ad hoc prescriptions, which lead to
vastly differing results for certain kinematics
\cite{deforest,koch,cdp}.

In this paper, we want to concentrate on particular classes of
non-\-rela\-tivis\-tic approximations for the current, and therefore
restrict our attention to the popular ansatz of applying the
electromagnetic current in its on-shell form. In subsequent work we
intend to widen the scope to include other approaches to more general
off-shell behaviour.

When calculating matrix elements for the single-nucleon case, the
initial and final states are plane waves, and therefore eigenfunctions
of the operator $\vec \eta$. In the nuclear case, the initial and
final states are single-particle wave functions for nucleons in the
nucleus. The coordinate space operator corresponding to $\vec \eta$ is
then $- i  \vec \nabla / \, m_N$.  Unlike the plane waves in the case
of the free nucleon, the single-particle wave functions are not
eigenfunctions of $\vec \eta$ or $\vec \nabla$, and this
leads to a technical problem, as the operator $\vec \eta$ appears
several times in the denominators of the $\xi$'s in
Eqs. (\ref{sn17}) and the $\nu$'s in
Eqs. (\ref{sn20}). Therefore, we have to Taylor-expand the expressions
for the current in powers of $\vec \eta$. Importantly, however, we
never make expansions in $\kappa$ or $\lambda$ (and therefore $\tau$),
as discussed below.

Equations (\ref{sn20}) have been cast in forms well suited for this
task.  We present the results to first order in $\eta$.  From
Eqs. (\ref{sn8}) and (\ref{sn13}) one finds that $\mu
_1=1+{\cal{O}}(\eta ^2)$ and $f_0=1+{\cal{O}}(\eta ^2)$, whereas one
sees that $\mu _2=1+\frac 12\sqrt{\frac \tau {1+\tau }}\eta
\cos \theta +{\cal{O}}(\eta ^2)$ by using
Eqs. (\ref{sn7}--\ref{sn9}). We then have
\begin{eqnarray}
\nu _0 &=&\frac \kappa {\sqrt{\tau }}G_E+{\cal{O}}(\eta ^2)  \nonumber \\
\nu _0^{\prime } &=&\frac \kappa {\sqrt{1+\tau }}\left[ G_M-\frac
12G_E\right] \eta \sin \theta +{\cal{O}}(\eta ^2)  \nonumber \\
\nu _1 &=&\frac 1{\sqrt{1+\tau }}\left[ G_E+\frac 12\tau G_M\right] \eta
\sin \theta +{\cal{O}}(\eta ^2)  \nonumber \\
\nu _1^{\prime } &=&\sqrt{\tau }G_M+{\cal{O}}(\eta ^2)  \label{sn28} \\
\nu _2^{\prime } &=&\sqrt{\tau }G_M+{\cal{O}}(\eta ^2)  \nonumber \\
\nu _2^{\prime \prime } &=&\frac 12\left( \frac \lambda \kappa \right) \sqrt{%
\tau }G_M\eta \sin \theta +{\cal{O}}(\eta ^2),  \nonumber
\end{eqnarray}
leading to the following results for the electromagnetic current
operators to linear order in $\eta $:
\begin{eqnarray}
\bar J^0 &=&\frac \kappa {\sqrt{\tau }}G_E+\frac i{\sqrt{1+\tau }}\left[
G_M-\frac 12G_E\right] \left( \vec{\kappa }\times \vec{\eta }\right)
\cdot \vec{\sigma }+{\cal{O}}(\eta ^2)  \label{sn29} \\
\bar J^3 &=&\left( \lambda /\kappa \right) \bar J^0  \label{sn30} \\
\vec{\bar J}^{\bot } &=&-\frac{\sqrt{\tau }}\kappa \left\{ iG_M\left(
\left[ \vec{\kappa }\times \vec{\sigma }\right] +\frac 12\left( \frac
\lambda \kappa \right) \frac 1\kappa \left( \vec{\kappa }\cdot \vec{%
\sigma }\right) \left( \vec{\kappa }\times \vec{\eta }\right) \right)
\right.  \nonumber \\
&&-\left. \left( G_E+\frac 12\tau G_M\right) \left[ \vec{\eta }-\left( 
\frac{\vec{\kappa }\cdot \vec{\eta }}{\kappa ^2}\right) \vec{\kappa 
}\right] \right\} +{\cal{O}}(\eta ^2),  \label{sn31}
\end{eqnarray}
employing Eqs. (\ref{sn14}--\ref{sn16}) and noting from
Eq. (\ref{sn7}) that $\kappa =\sqrt{\tau (1+\tau )}+\tau \eta \cos
\theta +{\cal{O}}(\eta ^2).$ Of course, when computing matrix
elements of these operators and then forming bilinear combinations of
the results to obtain the electromagnetic observables, terms of order
$\eta ^2$ must be neglected if the operators themselves have been
expanded only to order $\eta $, since other terms will enter from
considering the neglected ${\cal{O}}(\eta ^2)$ contributions in
Eqs. (\ref {sn29}--\ref{sn31}).

As can be seen from Eqs. (\ref{sn28}--\ref{sn30}), at linear order in
$\eta$ we retain the spin-orbit part of the charge and one of the
relativistic corrections to the transverse current that appeared
in Eq. (\ref{sn16}), the first order convective spin-orbit term.

The important point in our approach is that we have expanded only in
$\eta$, not in the transferred momentum $\kappa$ or the transferred
energy $\lambda$. The momentum of the initial nucleon will be
relatively low in most cases --- the dimensionless Fermi momentum
$\eta_F = p_F/m_N$ typically ranges from about $0.06$ for deuterium to
about $0.28$ for heavy nuclei. However, there is a lot of interest in
the investigation of the short--range properties of the nuclear wave
functions which are reflected in the behavior at large momentum, and
for those cases it is necessary to establish how good the
approximation in $\eta$ is and where it does not work.

In the next section, we will discuss the relativistic corrections to
the current for the special case of a deuterium target. The deuteron's
Fermi momentum of 55 MeV/c = 0.28 fm$^{-1}$, which corresponds to
$\eta_F \approx 0.06$, is considerably smaller than the Fermi momenta
of heavier nuclei.  For many applications, the initial momenta
involved are below the Fermi momentum, as this part of the wave
function leads to large and therefore experimentally more
accessible cross sections. Note that the Fermi momentum itself is not a
scale in the expansion in $\eta$, so that when we consider heavier
nuclei, the convergence at the respective higher Fermi momentum will
be worse than the convergence at the deuteron's smaller Fermi
momentum.  For this reason, we have extended our analysis of electron
scattering from the deuteron in the next section to rather high
missing momenta in order to draw conclusions about the
convergence of the expansion in $\eta$ also for the region $\eta >
\eta_F^{deuteron}$ which is accessible
experimentally for heavier nuclei due to their larger Fermi momenta.

Of course, for the high momentum components of the wave function,
off-shell effects are expected also to be important and for a complete
understanding of the problem have to be considered in addition to the
relativistic corrections discussed here.

We illustrate the importance of retaining the exact expressions for
$\kappa$, $\lambda$ and $\sqrt{\tau}$ with a few numbers: for a
momentum transfer $ q = 2 m_N$, i.e.  $\kappa = 1$, the transferred
energy under quasielastic conditions is $\omega \approx 1.2$ GeV
corresponding to $\lambda \approx 0.6$ and $\tau \approx 0.6$. This
leads to an extra factor $\kappa / \sqrt{\tau} \approx 1.3$ which
appears in the current matrix element, and which may enter squared in
the calculated observable, therefore leading to a $60 \%$ increase in
that observable.  Even for moderate momentum transfers, e.g. $ q =
m_N$, and thus $\kappa = 0.5$, the transferred energy is $\omega
\approx 400$ MeV, i.e.  $\lambda \approx 0.2$ and $\tau \approx 0.2$
and an extra factor $\kappa / \sqrt{\tau} \approx 1.1$, which still
gives rise to a $20 \%$ increase.  Apart from the factor $\kappa /
\sqrt{\tau}$ which originates from the product of the upper components
and the inverse factor $\sqrt{\tau} / \kappa $ which stems from the
product of upper and lower components of the Dirac spinors, there also
appear certain combinations of form factors in typical observables,
namely, one often finds $G_E^2$ together with $( \sqrt{\tau} \,
G_M)^2$.  The latter term would be neglected in a scheme that retains
only terms of leading order in $\sqrt{\tau}$, but in fact, for
example, for protons assuming dipole parameterizations for the form
factors those two terms become equal for $\tau \approx
0.13$, which corresponds to $|Q^2| \approx (0.98$ GeV/c $)^2$.  The
importance of the relativistic corrections will become even clearer in
the following section discussing the results for the reactions $^2 H
(e,e'p)n$ and $^2 \vec H (\vec e,e'p)n$.

As one can see from the expressions given above, it is actually
unnecessary to make expansions in $\kappa $, $\lambda $ or $\tau $ at
all. If one does so in spite of this, then at intermediate
momentum transfers the combination $G_M^{\prime }\equiv \sqrt{\tau
}G_M$ should be regarded as being of leading order, and not of order
$1 / m_N^{2}$ as is often assumed.  With these caveats under
some circumstances Eq. (\ref{sn31}) may be approximated by
\begin{equation}
\vec{\bar J}^{\bot }\cong -\frac{\sqrt{\tau }}\kappa \left\{ iG_M\left[ 
\vec{\kappa }\times \vec{\sigma }\right] -G_E\left[ \vec{\eta }%
-\left( \frac{\vec{\kappa }\cdot \vec{\eta }}{\kappa ^2}\right) 
\vec{\kappa }\right] \right\} .  \label{sn32}
\end{equation}

As we will compare our results with the strict non-relativistic reduction,
we quote the corresponding expressions:

\begin{eqnarray}
\bar J^o_{nonrel} & = & G_E \nonumber \\ 
\bar J^{\perp}_{nonrel} & = & -i G_M
\left [\vec \kappa \times \vec \sigma \right ] + G_E \left [ \vec \eta -
\left ( \frac{\vec \kappa \cdot \vec \eta}{\kappa^2} \right ) \vec \kappa
\right ] \,.
\label{curnr}
\end{eqnarray}
Note that the non-relativistic reduction contains both terms of zeroth
order in $\eta$ and terms of first order in $\eta$, i.e. the
convection current. The strict non-relativistic reduction is therefore 
not the lowest-order term of an expansion in $\eta$. 

In our new approach, once the on-shell form has been adopted, the only
approximation made is cutting off the expansion in powers of $\eta$ at
a certain order. Of course, it is necessary to check how good this
approximation is, e.g. if it is sufficient to use the expressions up
to first order in $\eta$, which are given explicitly above, or if it
is necessary to include higher orders. One possibility to check the
quality of the expansion is to evaluate the second-order expressions
in $\eta$ and compare them with the first-order results, but as we do
not know the convergence properties of the power series, this gives
only limited information.  It is clearly desirable to compare the
first-order results with the results obtained with the full form of
the operator. Due to the in general complicated form of the nuclear
initial and final states, this is not possible for most cases.
However, for coincidence $(e,e'N)$ reactions in Plane Wave Impulse
Approximation (PWIA), it is actually possible to carry out a
calculation with the full current operator.  In this special case, one
always has an outgoing plane wave for the knocked out nucleon in the
final state, and by partially integrating the single-particle matrix
elements of the current twice, the operator $\vec \eta \, ^2$ --- it
appears only squared, never in linear or cubic form --- acts on the
final state and thus on its eigenfunction.

\section{Results}

We have chosen to demonstrate the effects of the relativistic
corrections to the current for the reaction $^2 H (e,e'p)n$ and
$^2\vec H( \vec e, e'p)n$ because of the following reasons: first,
realistic wave functions are available in parameterized form for this
nucleus \cite{bonn,paris} and second, there is a lively experimental
interest in the exploration of the properties of few-body systems.
Specifically, there are several experiments planned at TJNAF with
deuterium targets \cite{cebaf}.  However, the current operator
discussed in the previous section can be applied to any nucleus and to
any type of electronuclear reaction.

An extensive discussion of coincidence reactions in general can
be found in \cite{raskintwd}; here we only quote the basic formulae.
The differential cross section is equal to

\begin{eqnarray}
\left ( \frac{ d \sigma^5}{d \epsilon' d \Omega_e d \Omega_N}
\right ) _{fi}^h  & = & 
\frac{m_N \, m_f \, p_N}{8 \pi^3 \, m_i} \, \sigma_{Mott} \, 
f_{rec}^{-1} \, \nonumber \\
& & \Big[ \left ( v_L {\cal{R}}_{fi}^L +   v_T {\cal{R}}^T_{fi}
 + v_{TT} {\cal{R}}_{fi}^{TT} + v_{TL} {\cal{R}}_{fi}^{TL} \right )
  \nonumber \\
& & +  h \left ( v_{T'} {\cal{R}}_{fi}^{T'} +  v_{TL'} {\cal{R}}_{fi}^{TL'}
\right ) \Big] \, ,
\end{eqnarray}
where $m_i$, $m_N$ and $m_f$ are the masses of the target nucleus, the
ejectile nucleon and the residual system, $p_N$ and $\Omega_N$ are the
momentum and solid angle of the ejectile, $\epsilon'$ is the energy of
the detected electron and $\Omega_e$ is its solid angle.  The helicity
of the electron is denoted by $h$.  The coefficients $v_K$ are the
leptonic coefficients, and the ${\cal{R}}_K$ are the response functions
which are defined by

\begin{eqnarray}
{\cal{R}}_{fi}^L & \equiv & | \rho (\vec q)_{fi}|^2 \nonumber \\
{\cal{R}}_{fi}^T & \equiv & | J_+ (\vec q)_{fi}|^2 
+ | J_- (\vec q)_{fi}|^2  \nonumber  \\
{\cal{R}}_{fi}^{TT} & \equiv &  2 \, \Re \, \big[ J_+^* (\vec q)_{fi} \,
J_- (\vec q)_{fi} \big] \nonumber  \\
{\cal{R}}_{fi}^{TL} & \equiv & - 2 \, \Re \, \big[ \rho^* (\vec q)_{fi} \,
( J_+ (\vec q)_{fi} - J_- (\vec q)_{fi}) \big] \nonumber \\
{\cal{R}}_{fi}^{T'} & \equiv & | J_+ (\vec q)_{fi}|^2 -
 | J_- (\vec q)_{fi}|^2  \nonumber  \\
{\cal{R}}_{fi}^{TL'} & \equiv & - 2 \, \Re \, \big[ \rho^* (\vec q)_{fi} \, 
( J_+ (\vec q)_{fi} + J_- (\vec q)_{fi}) \big] \, , 
\label{defresp}
\end{eqnarray}
where the $J_{\pm}$ are the spherical components of the current.

For our calculations, we have chosen the following kinematic
conditions: the z-axis is parallel to $\vec q$, the missing momentum
is defined as $\vec p_m \equiv \vec q - \vec p_N$, so that in PWIA,
the missing momentum is equal to the negative initial momentum of the
struck nucleon in the nucleus, $\vec p_m = -\vec p$. We denote the
angle between $\vec p_m$ and $\vec q$ by $\theta$, and the term
``parallel kinematics'' indicates $\theta = 0^o$, ``perpendicular
kinematics'' indicates $\theta = 90^o$, and ``antiparallel
kinematics'' indicates $\theta = 180^o$.  Note that both this
definition of the missing momentum and the definition with the other
sign are used in the literature.  If not stated otherwise, we assume
that the experimental conditions are such that the kinetic energy of
the outgoing nucleon and the angles of the missing momentum, $\theta$
and the azimuthal angle $\phi$, are fixed.  If not mentioned
otherwise, the kinetic energy of the outgoing proton is fixed to 1
GeV. For changing missing momentum, the transferred energy and
momentum change accordingly --- for the convenience of the reader, we
have listed the values of the transferred energy and momentum for
different kinematic conditions in Tables \ref{taben} and 
\ref{tabmom}.  In antiparallel kinematics, the transferred momentum
decreases with increasing missing momentum until a point is reached
where the kinematical limit for the process is reached, i.e. where the
transferred 4-momentum becomes timelike. This occurs at $p_m = 2.7$
fm$^{-1}$. For this reason, the curves showing results for
antiparallel kinematics are cut off at this value. The strong increase
or decrease that can be observed for some cases at this point is just
an artifact: when the responses are multiplied with the corresponding
leptonic coefficients, the product goes to zero.

The curves shown in this paper have been obtained using the realistic
Bonn wave function \cite{bonn} for the deuteron and
we do not present results for other wave functions
as we concentrate on the relativistic effects in the current, which
are expected to be similar no matter which wave function one employs.
In future work, other cases will be discussed.

\subsection{Unpolarized Responses}

In Fig. \ref{figrl}, we show the results for the longitudinal response
${\cal R}_L$ and in Fig. \ref{figrt} the transverse response ${\cal
R}_T$ calculated with the full current operator, the current operator
to ${\cal O} (\eta)$, and the strict non-relativistic limit.  In order
to facilitate the comparison between the different results in all
regimes of the missing momentum, we have included both linear and
logarithmic plots throughout this paper. For all the different
kinematics, the approximation of the electromagnetic current operator
to ${\cal O}(\eta)$ is in good agreement with the results obtained
with the full current. The two curves practically coincide up to
missing momenta $p_m \approx 1.3 - 1.5$ fm$^{-1}$, and the difference
for higher missing momentum is small. This behavior is to be expected,
as in PWIA the missing momentum coincides with the momentum of the
initial nucleon inside the nucleus, which is precisely the quantity in
which we have expanded the current operator. We have only expanded up
to the first order of the initial momentum, and it is quite surprising
that the agreement is so good up to such rather high momenta. This
indicates that the expansion coefficients of the higher powers of
$\eta$ are rather small.  In contrast to this, the results obtained
within the strict non-relativistic limit disagree strongly --- except
for parallel kinematics --- with the results that include relativistic
effects.  From the linear plots, it is clear that there is a
significant disagreement even for the low missing momenta. This
disagreement of relativistic and non-relativistic treatment of the
current increases up to almost one order of magnitude for the higher
$p_m$.

Let us examine the validity of the assumptions stated in
Eq. (\ref{nonrel}) that enter the strict non-relativistic reduction.
For the three different kinematics presented here, the transferred
energy $\omega$ increases with increasing missing momentum (see Table
\ref{taben}), and therefore the assumption that $\omega \ll m_N$ is
never fulfilled; it actually evolves to $\omega > m_N$ for higher
$p_m$.  The initial momentum, i.e. $p_m$, remains smaller than the
nucleon mass, but comes quite close to it.  Finally, the transferred
momentum $q$ behaves differently for the different kinematic
conditions (see Table \ref{tabmom}). For antiparallel kinematics, it
decreases; for perpendicular kinematics, it decreases slightly; and
for parallel kinematics, it increases. In all cases, however, it
remains above the nucleon mass.  For parallel kinematics, in spite of
the fact that both transferred energy and transferred momentum are
larger than $m_N$, the relation $\omega < q$ holds, which in the form
of $\omega \approx q^2 / 2 m_N$ enters the non-relativistic reduction.
This is the reason for the comparatively better performance of the
non-relativistic reduction in parallel kinematics.

After these general considerations, let us turn our attention to the
more specific effects of the relativistic corrections on the different
responses.  For the longitudinal response ${\cal R}_L$, the
relativistic corrections lead to an increase of the response
function. The bulk of this effects stems from the factor
$\kappa / \sqrt{\tau}$ which is contained in $\xi_0$ and
multiplies the zeroth-order charge operator. It reaches values
of more than $1.5$, and it enters squared in the response. In
addition, a new type of operator, the first-order spin-orbit term,
appears in the relativistic treatment.  This operator is already known
from the $q / m_N$ expansions, but in our treatment of the
relativistic effects it has a modified factor which multiplies
it. Note that in a $q / m_N$ expansion scheme, the factor
$\kappa / \sqrt{\tau}$ would be treated as $1$.

For the transverse response ${\cal R}_T$, the relativistic corrections
lead to a decrease of the response.  This is due to the factor
$\sqrt{\tau} / \kappa $ that multiplies the whole transverse current.
The fact that here $\sqrt{\tau}/ \kappa$ appears rather than its
inverse as in ${\cal R}_L$ stems from the fact that the latter arises
at leading order from upper $\times$ upper spinor components whereas
here upper $\times$ lower components occur at leading order, bringing
in an extra factor of $\tau / \kappa^2$.  The zeroth-order
magnetization current is by far the largest contribution to the
transverse current, and it completely dominates ${\cal R}_T$.  The
first-order convection current and the new types of operators which
appear only in the relativistic treatment become important only for
the interference response functions.

In order to get a different perspective, we present in
Fig.~\ref{rlrtyfig} the longitudinal response ${\cal R}_L$ (left side)
and the transverse response ${\cal R}_T$ (right side) for fixed
momentum transfer $|\vec q| = 1.4$ GeV/c, corresponding to $\kappa =
0.75$, and fixed $y$, which implies a fixed energy transfer in
turn. For the definition of $y$, see \cite{yscal}. The middle panels
of Fig.~\ref{rlrtyfig} show the responses under quasi-elastic
conditions, the top panels are kinematically ``below'' the
quasi-elastic peak, the lower panels are ``above'' the quasi-elastic
peak. In this choice of kinematics, the angle $\theta$ between the
missing momentum and the $z$-axis changes with the missing momentum,
as does the kinetic energy of the outgoing proton. In contrast to the
previously considered kinematic setting, the value of $\kappa /
\sqrt{\tau}$ does not change.  Again, there is good agreement for the
fully relativistic and first-order results.

The transverse response is dominated by the zeroth-order magnetization
current.  Therefore, the difference between the non-relativistic result
and the first-order result is given by the factor $\sqrt{\tau} /
\kappa$ which has the value $0.90$, $0.85$, and $0.74$ for the top,
middle, and lower panels.  Obviously, the relativistic effects are
largest for the largest energy transfer.  The exact
relativistic results tend to be larger than the first-order results at
higher missing momenta because,  although the
coefficient $\xi_1'$ of the zeroth-order magnetization current
decreases with increasing missing momentum, the overall factor $f_o$,
which only appears in the full calculation, increases.

In contrast to the transverse response, the longitudinal response has
significant contributions from two terms: the zeroth-order charge
operator and the first-order spin-orbit term.  The two contributions
do not interfere. The relativistic effects in
the zeroth-order charge operator are similar to the effects in the
zeroth-order magnetization current when one replaces the factor
$\sqrt{\tau} / \kappa$ with $\kappa /\sqrt{\tau} $.  The first-order
spin-orbit term does not enter in the non-relativistic calculation,
and its presence in the first-order results increases the difference
of the non-relativistic and first-order treatment.

In Fig. \ref{figint} , we show the interference response functions
${\cal R}_{TT}$ and ${\cal R}_{TL}$. The transverse-transverse
response function is negative throughout the considered range, and is
very small relative to ${\cal R}_{L,T}$. As discussed in the previous
section (see Eqs.(\ref{sn24}-\ref{sn25})) this behavior is expected,
since the spin-averaged unpolarized response is proportional to
$\delta^2$.  As these responses both vanish in parallel and
antiparallel kinematics, we present them only for perpendicular
kinematics.  The transverse-longitudinal response function falls
between ${\cal R}_{L,T}$ and ${\cal R}_{TT}$ in magnitude, again as
expected from the guidance provided by the spin-averaged unpolarized
response discussed above (see Eq.(\ref{sn25a})) where this
contribution is seen to be proportional to $\delta$.  Here the full
results and the results to first order in the initial nucleon momentum
agree extremely well up to $p_m = 4$ fm$^{-1}$. The non-relativistic
results are considerably lower over the whole range of missing
momenta, roughly by a factor $5$.  At first sight, this may look
surprising, as one might expect that the factors which were
responsible for the main difference between the relativistic and
non-relativistic results for the longitudinal and the transverse
responses, namely $\kappa / \sqrt{\tau}$ and $\sqrt{\tau} / \kappa$
would be multiplied yielding $1$, and that therefore there would be no
overall relativistic effects.  However, the transverse-longitudinal
response function has a different structure than ${\cal R}_L$ and
${\cal R}_T$, as seen in Eq. (\ref{defresp}).  The response ${\cal
R}_{TL}$ consists of two different contributions: one contains the
product of the first-order spin-orbit term and the zeroth-order
magnetization current, the other one contains the product of the
zeroth-order charge operator and the first-order convection current
(see also \cite{quiqueinc,quiquecoin}). The former amounts to roughly
two thirds of the total response, the latter to one third. As the
spin-orbit operator appears only in the relativistic treatment, it is
clear that the major contribution to ${\cal R}_{TL}$ is completely
missed in the strict non-relativistic limit. Also, the second
contribution increases when the relativistic effects are taken into
account: besides the two factors mentioned above which cancel each
other, the factor that multiplies the first-order convection current
increases from $G_E$ to $G_E + \frac{1}{2} \, \tau \, G_M$, which for
protons gives an extra factor of approximately $ 1 + 1.4 \tau$,
assuming dipole parameterization for the form factors.  Note that
those results hold under all kinematical conditions --- they are not
specific for perpendicular kinematics. In \cite{tjon}, a microscopic
calculation was carried out for the $^2H(e,e'p)$ reaction at lower
energies, and this calculation found large relativistic effects in
${\cal R}_{TL}$. As our results show, those large relativistic effects
can be attributed to the current; it is not necessary to incorporate
relativistic dynamics in order to see large relativistic effects 
(see also \cite{pd}).

As noted above, the transverse-transverse response function is the
smallest of all the responses. As can be seen in Fig. \ref{figint},
for this response the relative difference between the first-order and
full results is the largest one we have encountered so far.  This is
to be expected, as the transverse-transverse response is explicitly a
quantity of second order in $\eta$ (see Eqs.(\ref{sn24}-\ref{sn25})),
so the second-order contributions to the current, i.e. the
second-order convective spin-orbit term, will contribute significantly
to this response, too.  Up to $p_m = 1.5$ fm$^{-1}$, the first-order
results are slightly larger than the non-relativistic results.  For
$p_m > 3$ fm$^{-1}$, the first-order result starts to be smaller and
disagrees more with the exact result, whereas the non-relativistic and
exact relativistic calculations approach each other.  The
transverse-transverse response function is the product of two currents
and because of this it has a unique structure, as seen in
Eq. (\ref{defresp}). Due to this structure, the zeroth-order
magnetization current by itself cannot contribute.  This fact reflects
that the response is of second order --- it therefore cannot have a
zeroth-order contribution.  As the zeroth-order magnetization current
is the largest of the components of the transverse current, this
explains why ${\cal R}_{TT}$ is so small.  It also means that in the
non-relativistic limit, only the first-order convection current can
contribute.  At first order in the expansion in $\eta$, the
first-order convective spin-orbit term (see Eq. (\ref{sn30})),
enters. This relativistic correction gives rise to a new, additive
contribution to ${\cal R}_{TT}$. It has the opposite sign compared
with the first-order convection current contribution, and although it
is small, at higher $p_m$ it is big enough to overcompensate the rise
in the first-order convection current contribution due to relativistic
effects.  From the first order to the full current, the first-order
convection current contribution rises more than the relativistic
correction contribution, which in itself already amounts to an
increase in ${\cal R}_{TT}$. However, there is still another new
component of the current that arises only in the exact expression, the
second-order convective spin-orbit term (see Eq.~(\ref{sn16})). The
contribution of this new operator alone is quite small, but due to the
different spin structure that it has (the same as the first-order
spin-orbit term in the charge operator), it can interfere with the
zeroth-order magnetization current, which leads to a sizable new
contribution of second order in $\eta$. This new contribution is the
main reason for the discrepancy between the full result and the
first-order result: one of the major contributions to the
transverse-transverse response function is present only in the exact
treatment of the current.  In the full relativistic calculation, the
second-order convective spin-orbit term also interferes with the
first-order convection current and the first-order convective
spin-orbit term, but those contributions are much smaller than that of
the interference of the second-order convective spin-orbit term with
the zeroth-order magnetization current.  The fact that the
non-relativistic results seem to approach the exact ones for higher
missing momentum is therefore mere coincidence.

In Fig.~\ref{intyfig}, we show the interference response functions
${\cal R}_{TT}$ and ${\cal R}_{TL}$ for the same kinematic conditions
as used in Fig.~\ref{rlrtyfig}. For the transverse-longitudinal
response, the value of $\kappa /\sqrt{\tau} $ is not significant
because of the cancellation discussed earlier, and therefore the
difference between the non-relativistic and relativistic results is
roughly the same for all different kinematics, in contrast to the
longitudinal and transverse responses.  For the transverse-transverse
response, the difference between the non-relativistic and exact
relativistic results also remains roughly constant under the different
kinematic conditions, but the first-order results shift from the
vicinity of the exact results towards the non-relativistic curve, and
for the highest energy transfer, coincides more or less with the
non-relativistic results. This shows that terms of second order in
$\eta$ play an important role here, namely the second-order convective
spin-orbit term, which was discussed above.

\subsection{Polarized Responses}

After this discussion of the response functions that arise for
unpolarized beam and unpolarized target nucleus, let us discuss some
of the polarization observables. As the number of polarization
response functions is quite high, we will just discuss two
representative examples: the response ${\cal R}_{T'}^{M_J=1}$, which
is nonzero only for a polarized target and polarized electron beam,
and the response ${\cal R}_{TL'}^{M_J=1}$.  For the discussion of
those responses, we will assume that the target is completely
polarized in the $M_J = 1$ state, hence the superscript $M_J = 1$. The
direction of the momentum transfer serves as polarization axis.

In Fig.~{\ref{fig3}, the absolute value of the response ${\cal
R}_{T'}^{M_J=1}$ is shown for calculations with the full current
operator, the current operator to ${\cal O} (\eta)$, and the strict
non-relativistic limit.  The behavior of the different curves is very
similar to the case of the transverse response function discussed
previously: For antiparallel and perpendicular kinematics, the
non-relativistic result differs from the relativistic treatment
considerably, especially for high $p_m$.  For parallel kinematics, all
curves are quite similar, for the reasons discussed above.  The
agreement between the first-order calculation and the full
calculation is excellent at lower missing momenta.  The largest
contribution to ${\cal R}_{T'}^{M_J=1}$ comes from the zeroth-order
magnetization current, and its behavior governs the behavior of the
response. The relativistic treatment introduces the factor
$\sqrt{\tau} / \kappa $, which reduces the current and the
response. In the non-relativistic limit, the contribution from the
first-order convection current is negligible. To first order in
$\eta$, there is a very small contribution from the interference of
the first-order convection current and the first-order convective
spin-orbit term. In the exact calculation, there is an additional
contribution from the interference between the zeroth-order
magnetization current and the second-order convective spin-orbit term
which appears only in the full treatment of the current
operator. However, those are only small corrections to the dominant
contribution of the zeroth-order magnetization current itself.

For in-plane kinematics, the ${\cal R}_{TL'}^{M_J=1}$ vanishes in
parallel and antiparallel kinematics, so we show it only for
perpendicular kinematics in Fig.~\ref{fig4}. The most striking feature
of Fig.~\ref{fig4} is the fact that in the non-relativistic limit,
this response is negligible; the actual numbers are around $10^{-14}$
fm$^{-3}$ sr$^{-1}$, whereas the response is quite sizable once the
relativistic corrections are included. For the special case of
perpendicular kinematics at the non-relativistic level, the only
contribution to ${\cal R}_{TL'}^{M_J=1}$ contains the product of the
zeroth-order charge operator and the first-order convection
current. For general kinematics (see the discussion of the next
figure), the product of the zeroth-order charge operator and the
zeroth-order magnetization current also enters.  Once the first order
in $\eta$ is considered, the products of the zeroth-order
magnetization current and the first-order spin-orbit term of the
charge operator and the product of the first-order convective
spin-orbit term and the zeroth-order charge operator contribute to the
response, leading to a magnitude that is comparable with the magnitude
of ${\cal R}_{TL}$. Again, the contribution that includes the
zeroth-order magnetization current is the dominant one. The second
contribution has the opposite sign and its size is about $20 \%$ to
$30 \%$ of the zeroth-order magnetization current/first-order
spin-orbit contribution.  For first order in $\eta$, the response is
negative for low $p_m$, it changes sign at $p_m = 1.6$ fm$^{-1}$ and
continues to be positive.  The full result coincides with the
first-order results up to $p_m \approx 2 $ fm $^{-1}$, but it later
changes sign again, therefore becoming quite different from the
first-order result for the highest missing momenta considered here.

In Fig.~\ref{rtlpyfig}, we show the response ${\cal R}_{TL'}^{M_J=1}$
for fixed $q$ and $y$. Under those non-perpendicular conditions, the
non-relativistic result is of comparable size to that of the relativistic
result. The vanishing of the product of the zeroth-order charge
operator and the zeroth-order magnetization current in perpendicular
kinematics is due to the fact that it is proportional to spherical
harmonics that vanish for $\theta = 90^o$.  Note that the spherical
harmonics would be modified (and non-zero) once a final state
interaction is included.  Still, the difference between
non-relativistic and relativistic treatment is considerable, especially
for the higher values of $y$ and therefore of the energy transfer.

The current presented in this paper can be applied to heavy nuclei
immediately, and the relativistic effects for heavy nuclei in a
certain kinematical situation, i.e.  for a given set of $p_m$,
$\theta$, and $q$ and $\omega$, will be similar to what has been found
for the deuteron.  One should keep in mind that the Fermi momentum of
heavy nuclei is larger than the rather small Fermi momentum of the
deuteron. Thus, even if one takes the point of view that one is
interested only in the region below the Fermi momentum, where the
cross section is larger and consequently easier to access
experimentally than at high $p_m$, one will need to consider higher
missing momenta for heavier nuclei where relativistic effects are
expected to be larger.

\section{Conclusions}

In this paper, we have investigated the relativistic effects that
appear in the electromagnetic current operator.  In order to perform a
calculation of electronuclear reactions, we have chosen the popular
off-shell prescription of using what we have called the ``on-shell form
of the operator''.

There are several calculations of particular electronuclear reactions for
few-body targets in a microscopic fashion \cite{tjon,vano}.
Naturally, these approaches contain relativistic treatments
of the nuclear dynamics and the current. However, it is difficult to
extend these approaches to higher energies --- the problems start
above the pion emission threshold --- and they appear unlikely to be
applied to heavier nuclei in the near future.  The current operator
discussed in this paper is not restricted by these limitations, as it
can be used regardless of energy and target.

We have presented a formalism that allows one to perform the
non-relativistic reduction of the free single-nucleon electromagnetic
current operator without any approximation in the transferred momentum
or transferred energy. Within the context of the chosen so-called
on-shell form
the only remaining approximation is to
expand in powers of the initial momentum of the nucleon.  In
this paper, we have carried out a systematic investigation of the
quality of this approximation by comparing the results obtained with
the current to first order in $\eta$ with the full results for the
special case of the $(e,e'p)$ coincidence reaction in Plane Wave
Impulse Approximation. This is the only case for which the exact
relativistic calculation of the current can be applied to a realistic
nuclear wave function. Although the assumption of PWIA at those
energies is incomplete (see e.g. \cite{Deuter,Tensor}), the PWIA is
a valuable and reliable testing ground for the relativistic effects in
the operator.

We have found very good agreement between the full relativistic
treatment and the first-order results in almost all cases.  In the
special case of the PWIA, the quantity in which we have expanded, the
momentum of the initial nucleon, coincides with the negative missing
momentum.  As expected, the ${\cal O} (\eta)$ and exact results may
become somewhat different for missing momenta $p_m \approx 3$ fm
$^{-1}$ and higher.

Naturally, the size of the relativistic corrections depends on the
specific kinematics. The
non-relativistic results are far from the full calculation and even far
from the first-order results, especially for non-parallel kinematics.

Our results show that at GeV energies, it is necessary to take into
account the relativistic corrections to the electromagnetic current
operator.  We have also shown that it is satisfactory to include these
corrections up to the first order in the initial momentum of the
nucleon.  The results in this paper have been obtained for the
reactions $^2H(e,e'p)n$ and $^2 \vec H (\vec e, e'p)n$, but the
conclusions drawn here can immediately be applied to other target
nuclei and, using straightforward extensions of the ideas presented,
to all other electronuclear reactions.

\acknowledgments 
The authors thank J. E. Amaro for many useful discussions.
S.J.  is grateful to the Alexander von
Humboldt Foundation for the financial support she receives as a
Feodor-Lynen Fellow. This work was in part supported by
funds provided by the U.S. Department of Energy (D.O.E.)
under cooperative research agreement \#DF-FC02-94ER40818.


\begin{table}
\caption{ The transferred energy $\omega$ in GeV and the dimensionless
energy transfer $\lambda$ for increasing missing momentum $p_m$.
The kinetic energy of the outgoing proton is 1 GeV. The energy transfer
does not depend on the angles of the missing momentum.}
\label{taben}
\begin{tabular}{ccc}
$p_m$/fm$^{-1}$ & $\omega$/ GeV & $\lambda$ \\
\tableline 
0 & 1.00 & 0.53 \\
1 & 1.02 & 0.55 \\
2 & 1.08 & 0.58 \\
3 & 1.17 & 0.63 \\ 
\end{tabular}
\end{table}

\begin{table}
\caption{ The transferred momentum $q$ in GeV/c, the dimensionless
momentum transfer $\kappa$, the dimensionless negative 4-momentum
transfer $\tau$, and the ratio $\kappa$ / $\protect {\sqrt{\tau}}$ for
increasing missing momentum $p_m$.  The kinetic energy of the outgoing
proton is 1 GeV. The momentum transfer depends on the angles of the
missing momentum: the top part gives the values for parallel
kinematics, the middle part for perpendicular kinematics, and the
lower part for antiparallel kinematics. Note that for parallel
kinematics, the highest accessible missing momentum is $p_m = 2.7$
fm$^{-1}$.}
\label{tabmom}
\begin{tabular}{ccccc}
$p_m$/fm$^{-1}$ & $q$ / $\, \, $GeV/c & $\kappa$ & $\tau$ & 
$\kappa / \sqrt{\tau}$ \\ [0.5ex]
\tableline
0 & 1.70 & 0.90 & 0.53 & 1.23\\
1 & 1.89 & 1.01 & 0.72 & 1.19\\
2 & 2.09 & 1.11 & 0.91 & 1.16\\
3 & 2.29 & 1.22 & 1.10 & 1.16\\ 
\hline 
0 & 1.70 & 0.90 & 0.53 & 1.23\\
1 & 1.68 & 0.90 & 0.51 & 1.26\\
2 & 1.65 & 0.88 & 0.44 & 1.33\\
3 & 1.59 & 0.85 & 0.33 & 1.48\\
\hline 
0 & 1.70 & 0.90 & 0.53 & 1.23\\
1 & 1.50 & 0.80 & 0.34 & 1.37\\
2 & 1.30 & 0.69 & 0.15 & 1.78\\
3 & 1.10 & 0.59 & $<$ 0 & - \\ 
\end{tabular}
\end{table}

\begin{figure}
\caption {The longitudinal response function ${\cal R}_L$ is shown for
parallel kinematics (a and b), perpendicular kinematics (c and d), 
and antiparallel kinematics (e and f). The solid line shows the
response calculated with the full expression for the current
operator, the dashed line is the result of the ${\cal O} (\eta)$
calculation and the dash-dotted line represents the result of the
strict non-relativistic reduction. Note the different ranges of the
missing momentum $p_m$ for the linear plots (left) and the logarithmic
plots (right). }
\label{figrl}
\end{figure}

\begin{figure}
\caption {The transverse response function ${\cal R}_T$ is shown for
parallel kinematics (a and b), perpendicular kinematics (c and d), and
antiparallel kinematics (e and f). The solid line shows the response
calculated with the full expression for the current operator, the
dashed line is the result of the ${\cal O} (\eta)$ calculation and the
dash-dotted line represents the result of the strict non-relativistic
reduction. Note the different ranges of the missing momentum $p_m$ for
the linear plots (a,c,e) and the logarithmic plots (b,d,f).  }
\label{figrt}
\end{figure}

\begin{figure}
\caption {The longitudinal response function ${\cal R}_L$ (a,c,e) and
the transverse response function ${\cal R}_T$ (b,d,f) are shown for
fixed 3-momentum transfer $|\vec q|$ and different values of the $y$
variable, which is defined as the negative minimal missing momentum
(see e.g.{\protect \cite{yscal}}).  The corresponding fixed energy
transfers $\omega$ are 0.61 GeV (a and b), 0.75 GeV (c and d),
and 0.94 GeV (e and f).  The kinematic conditions in the
middle panels correspond to quasifree conditions. Note that the angle
$\theta$ varies with changing missing momentum. The solid line shows the
response calculated with the full expression for the current
operator, the dashed line is the result of the ${\cal O} (\eta)$
calculation and the dash-dotted line represents the result of the
strict non-relativistic reduction.}
\label{rlrtyfig}
\end{figure}

\begin{figure}
\caption {The negative transverse-transverse response
function ${\cal R}_{TT}$ (a and b) and the transverse-longitudinal
response function ${\cal R}_{TL}$ (c and d) are shown for
perpendicular kinematics.  The solid line shows the response
calculated with the full expression for the current operator, the
dashed line is the result of the ${\cal O} (\eta)$ calculation and the
dash-dotted line represents the result of the strict non-relativistic
reduction. Note the different ranges of the missing momentum $p_m$ for
the linear plots (a,c,e) and the logarithmic plots (b,d,f). }
\label{figint}
\end{figure}

\begin{figure}
\caption {The negative transverse-transverse response
function ${\cal R}_{TT}$ and the transverse-longitudinal response
function ${\cal R}_{TL}$ are shown for fixed 3-momentum transfer
$|\vec q|$ and different values of the $y$ variable.  The
corresponding fixed energy transfers $\omega$ are 0.61 GeV (a
and b), 0.75 GeV (c and d), and 0.94 GeV (e and f).  The
kinematic conditions in the middle panels correspond to quasifree
conditions. Note that the angle $\theta$ varies with changing
missing momentum. The solid line shows the
response calculated with the full expression for the current
operator, the dashed line is the result of the ${\cal O} (\eta)$
calculation and the dash-dotted line represents the result of the
strict non-relativistic reduction. }
\label{intyfig}
\end{figure}

\begin{figure}
\caption {The absolute value of the response function ${\cal
R}_{T'}^{M_J=1}$ is shown for parallel kinematics (a and b),
perpendicular kinematics (c and d), and antiparallel kinematics (e and f).
The response is negative for parallel and antiparallel kinematics, for
perpendicular kinematics it starts out negative and switches sign at
$p_m = 1.6$ fm $^{-1}$.  The solid line shows the response calculated
with the full expression for the current operator, the dashed line is
the result of the ${\cal O} (\eta)$ calculation and the dash-dotted
line represents the result of the strict non-relativistic
reduction. The target is assumed to be completely polarized in the
$M_J = 1$ state.  Note the different ranges of the missing momentum
$p_m$ for the linear plots (a,c,e) and the logarithmic plots (b,d,f).}
\label{fig3}
\end{figure}

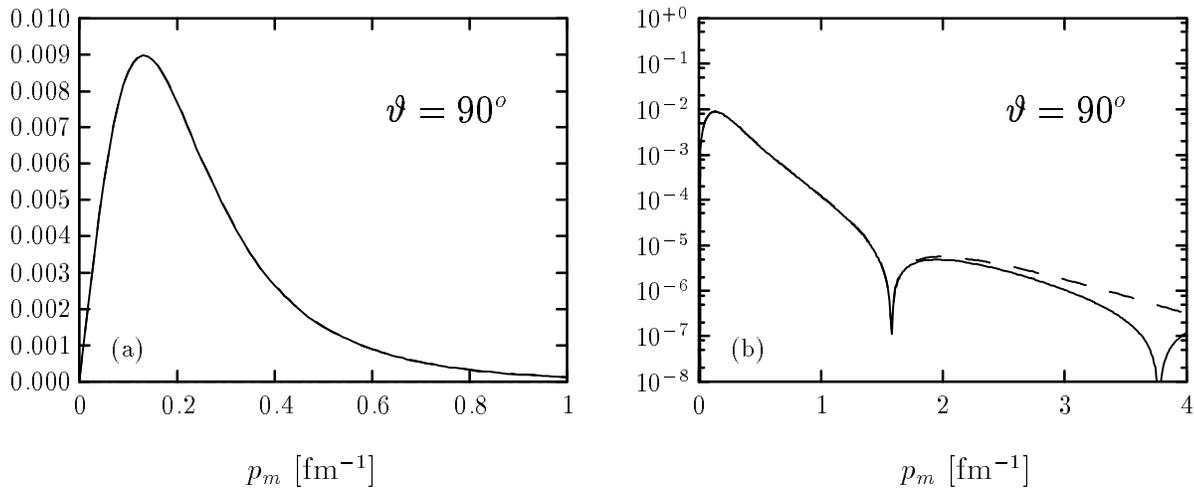
\begin{figure}
\caption {The absolute value of the response function ${\cal
R}_{TL'}^{M_J=1}$ is shown for perpendicular kinematics. The solid
line shows the response calculated with the full expression for the
current operator, the dashed line is the result of the ${\cal O}
(\eta)$ calculation and the dash-dotted line represents the result of
the strict non-relativistic reduction. Both first order and exact
result are negative for low $p_m$ and change sign at $p_m = 1.6$ fm
$^{-1}$. Whereas the first order result continues to be positive, the
exact result changes sign again for $p_m = 3.8$ fm $^{-1}$.  The
target is assumed to be completely polarized in the $M_J = 1$
state. Note the different ranges of the missing momentum $p_m$ for the
linear plot (a) and the logarithmic plot (b). }
\label{fig4}
\end{figure}

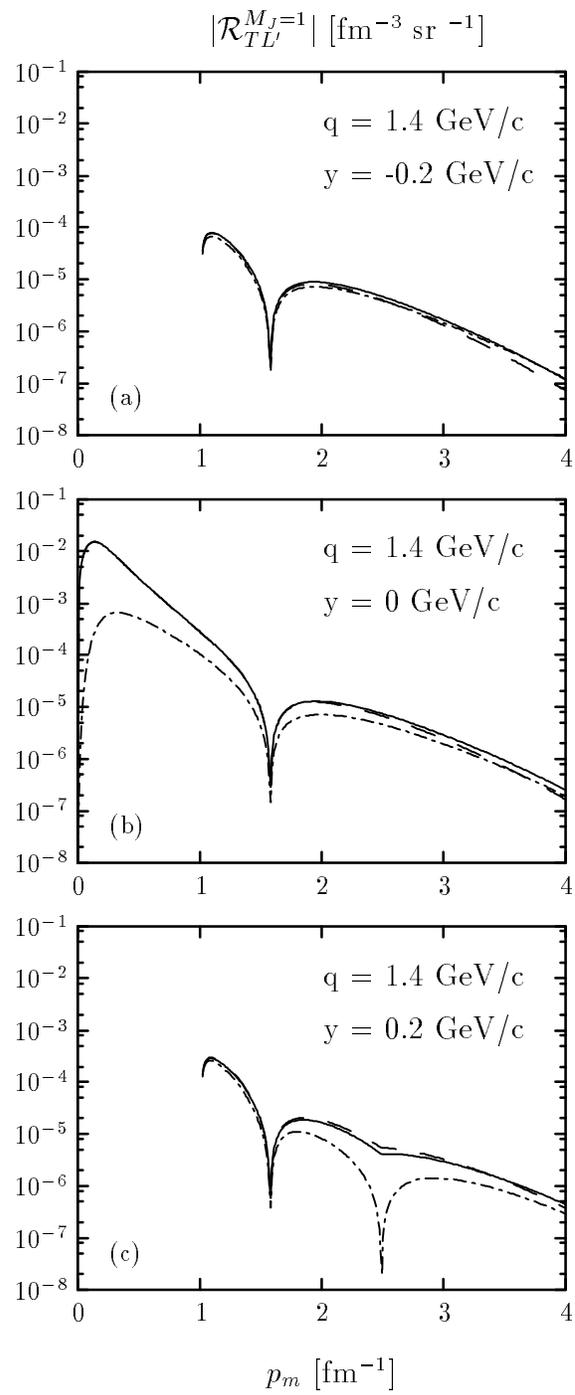
\begin{figure}
\caption {The absolute value of the response function ${\cal
R}_{TL'}^{M_J=1}$ is shown for fixed 3-momentum transfer $|\vec q|$
and different values of the $y$ variable.  The corresponding fixed
energy transfers $\omega$ are 0.61 GeV (a), 0.75 GeV (b),
and 0.94 GeV (c).  The kinematic conditions in the
middle panels correspond to quasifree conditions. Note that the angle
$\theta$ varies with changing missing momentum. The solid line shows the
response calculated with the full expression for the current
operator, the dashed line is the result of the ${\cal O} (\eta)$
calculation and the dash-dotted line represents the result of the
strict non-relativistic reduction. In all cases, the
response has a minus sign up to $p_m = 1.5$ fm$^{-1}$, and becomes and
stays positive afterwards.  }
\label{rtlpyfig}
\end{figure}

\end{document}